\documentclass[graybox,natbib,nosecnum]{svmult}
\bibpunct{(}{)}{;}{a}{}{,} 

\pdfoutput=1   

\usepackage{mathptmx}       
\usepackage{helvet}         
\usepackage{courier}        
\usepackage{type1cm}        

\usepackage{makeidx}         
\usepackage{graphicx}        
\usepackage{multicol}        
\usepackage[bottom]{footmisc}
\usepackage[normalem]{ulem}	
\usepackage{hyperref}  
\usepackage{soul}   
\usepackage[colorinlistoftodos,textsize=small]{todonotes} 

\newcommand{\hbindex}[1]{\hl{#1}\index{#1}}  

\makeindex             

\begin{document}

\title*{Accretion of Planetary Material onto Host Stars}
\author{Brian Jackson and Joleen Carlberg}
\institute{Brian Jackson \at Boise State University, Dept. of Physics, 1910 University Drive, Boise, Idaho, USA 83725-1570, \email{bjackson@boisestate.edu}
\and Joleen Carlberg \at Space Telescope Science Institute \email{jcarlberg@stsci.edu}}

\maketitle

\abstract{Accretion of planetary material onto host stars may occur throughout a star's life. Especially prone to accretion, extrasolar planets in short-period orbits, while relatively rare, constitute a significant fraction of the known population, and these planets are subject to dynamical and atmospheric influences that can drive significant mass loss. Theoretical models frame expectations regarding the rates and extent of this planetary accretion. For instance, tidal interactions between planets and stars may drive complete orbital decay during the main sequence. Many planets that survive their stars' main sequence lifetime will still be engulfed when the host stars become red giant stars. There is some observational evidence supporting these predictions, such as a dearth of close-in planets around fast stellar rotators, which is consistent with tidal spin-up and planet accretion. There remains no clear chemical evidence for pollution of the atmospheres of main sequence or red giant stars by planetary materials, but a wealth of evidence points to active accretion by white dwarfs. In this article, we review the current understanding of accretion of planetary material, from the pre- to the post-main sequence and beyond. The review begins with the astrophysical framework for that process and then considers accretion during various phases of a host star's life, during which the details of accretion vary, and the observational evidence for accretion during these phases.}

\section{Where are Planetary Bodies Disrupted?}
\label{sec:Where_Disrupted}
\begin{figure}
\includegraphics[width=\textwidth]{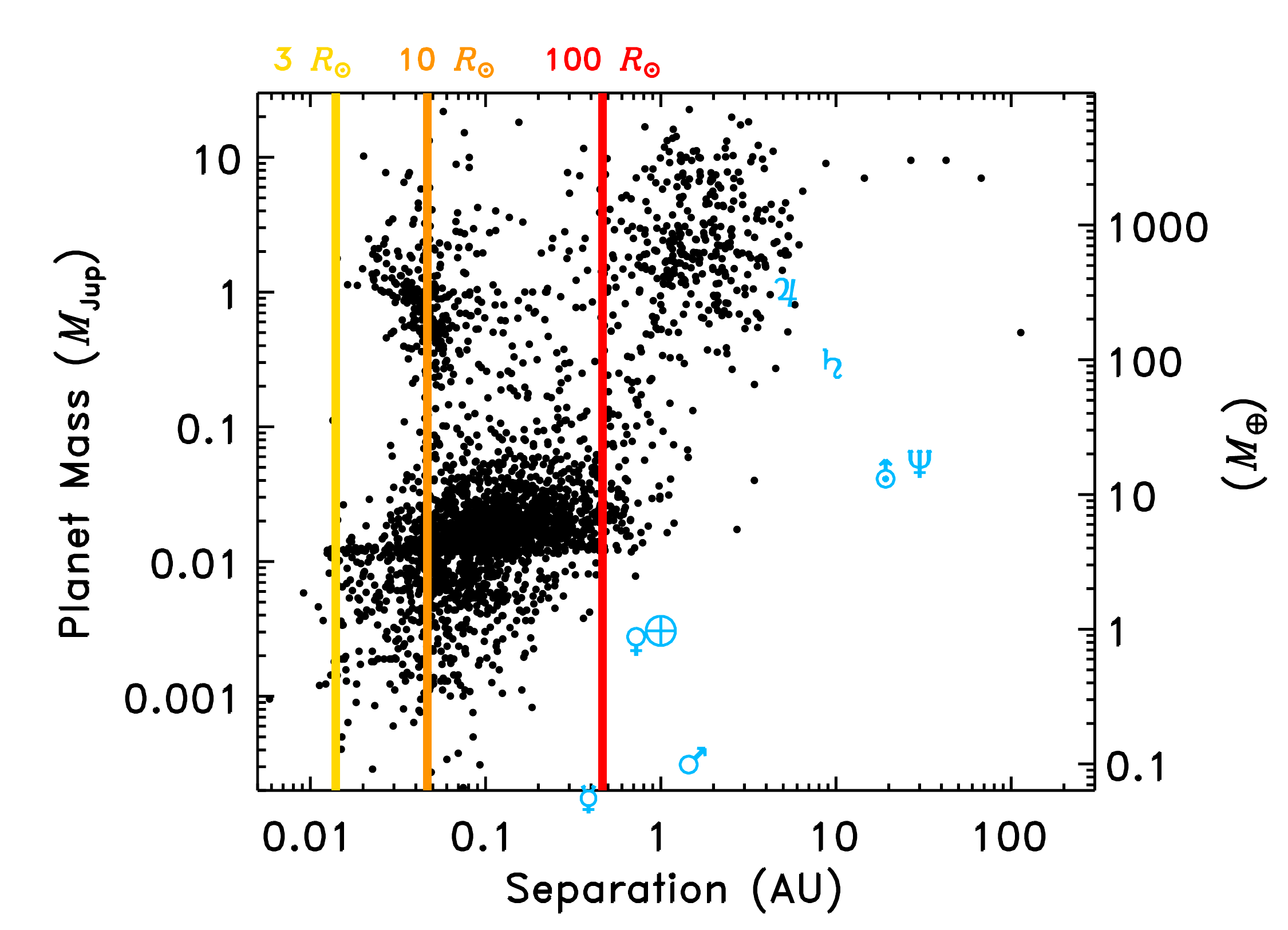}
\caption{Masses and separation of known exoplanets \citep{2014PASP..126..827H}. Blue symbols denote Solar System planets. Vertical lines show the size of the stellar radius at various stages of the post-main sequence for a Sun-like star.}
 \label{fig:Jackson-Carlberg_fig1}
\end{figure}

Owing to detection biases, exoplanet discoveries skew heavily toward short-period or close-in planets, as shown in Figure \ref{fig:Jackson-Carlberg_fig1}, and the nearest a planet can orbit its star and still remain intact depends largely on its self-gravity and, for smaller bodies, its tensile strength, friction, and/or internal viscosity. The \hbindex{Roche potential} plays a key role either way, and a considerable literature has developed around the Roche potential \citep{1959cbs..book.....K, 1971ARA&A...9..183P, 1993ApJS...88..205L, 1999ssd..book.....M}.

\begin{figure}
\includegraphics[width=\textwidth]{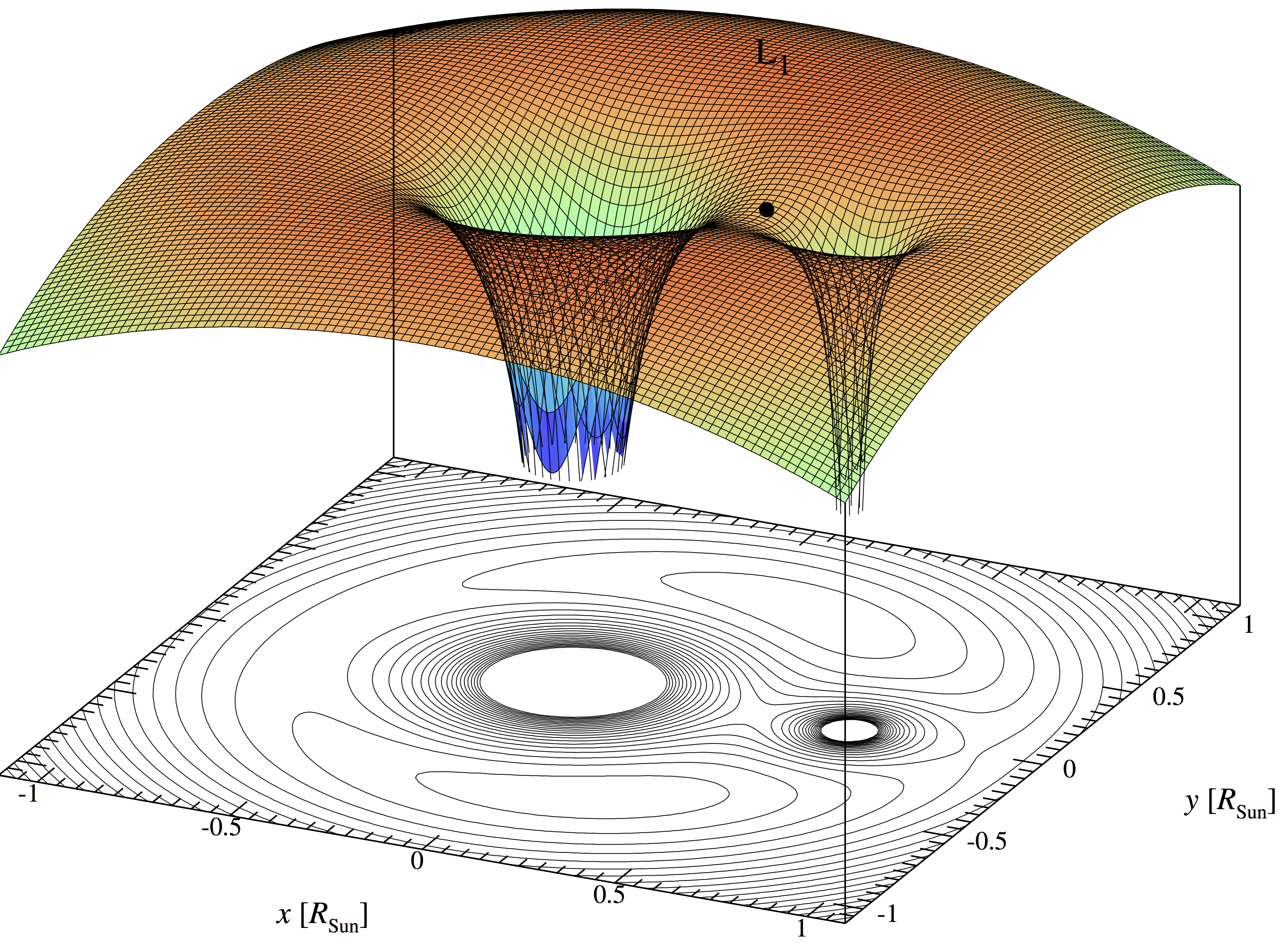}
\caption{The Roche potential. The colored grid shows a three-dimensional representation of the potential surface in the $x$-$y$ plane. Contours at the bottom show a projection of isopotentials onto the $x$-$y$ plane. The Lagrange point L1 is labeled. For this figure, $q = 0.2$ and distances are measured in solar radii $R_{\rm Sun}$, with the two masses separated by $0.718\ R_{\rm Sun}$. Courtesy of Ren\'{e} Heller.}
\label{fig:Jackson-Carlberg_fig2}
\end{figure}

The Roche potential is the potential field around a gravitating binary system in a bound orbit (Figure \ref{fig:Jackson-Carlberg_fig2}) and is cast in a frame that rotates with the binary and so includes a centrifugal term. The potential is dominated by the planet's gravity near the planet, but moving radially outward, the planet's gravity drops off until the potential reaches a local maximum, which corresponds to zero net acceleration. The corresponding surface around the planet is defined as the \hbindex{Roche lobe} \citep{1999ssd..book.....M}. Such a surface also surrounds the star.

The coordinate reference frame revolves with the system and is centered on the planet with the $x$-axis pointing from the planet to the star (which are separated by a fixed orbital distance $a$) and the $z$-axis pointing parallel to the orbital angular momentum. Approximating the two bodies as point masses, the effective potential (gravitational + centrifugal) field $\Phi$ is
\begin{equation}
\Phi = - \frac{GM_{\rm p}}{|\vec{r}|} - \frac{GM_{\rm s}}{| a \hat{x} - \vec{r} |} - \frac{1}{2} \Omega^2 \left[ \left(x - x_{\rm cm}\right)^2 + y^2 \right],
\label{eq:phi_full}
\end{equation}
where $G$ is the gravitational constant, $M_{\rm s}/M_{\rm p}$ is the stellar/planetary mass, $\vec{r}$ the location at which to evaluate $\Phi$, and $\hat{x}$ a unit vector pointing along $x$. The system's center of mass is at $x_{\rm cm} = a \left[ M_{\rm s}/\left( M_{\rm s}+M_{\rm p} \right) \right]$. $\Omega$ is the orbital mean motion, and $\Omega^2 = G \left( M_{\rm p}+M_{\rm s} \right) /a^3.$  

The Roche lobe is not spherical, and there is no general closed form expression for that surface. However, \citet{1983ApJ...268..368E} provided a formula for the volume-equivalent radius of the Roche lobe $R_{\rm RL}$, i.e., the radius of a sphere with a volume equivalent to the given Roche lobe's, accurate to 1\% for all mass ratios $q = M_{\rm p}/M_{\rm s}$:
\begin{equation}
R_{\rm RL} = \frac{0.49\ q^{2/3}}{0.6\ q^{2/3} + \ln\left( 1 + q^{1/3} \right)} a,
\end{equation}
and for $q \ll 1$
\begin{equation}
R_{\rm RL} \approx 0.49\ q^{1/3}\ a.
\label{eq:small_q_roche_lobe}
\end{equation}

\begin{figure}
\includegraphics[width=0.95\textwidth]{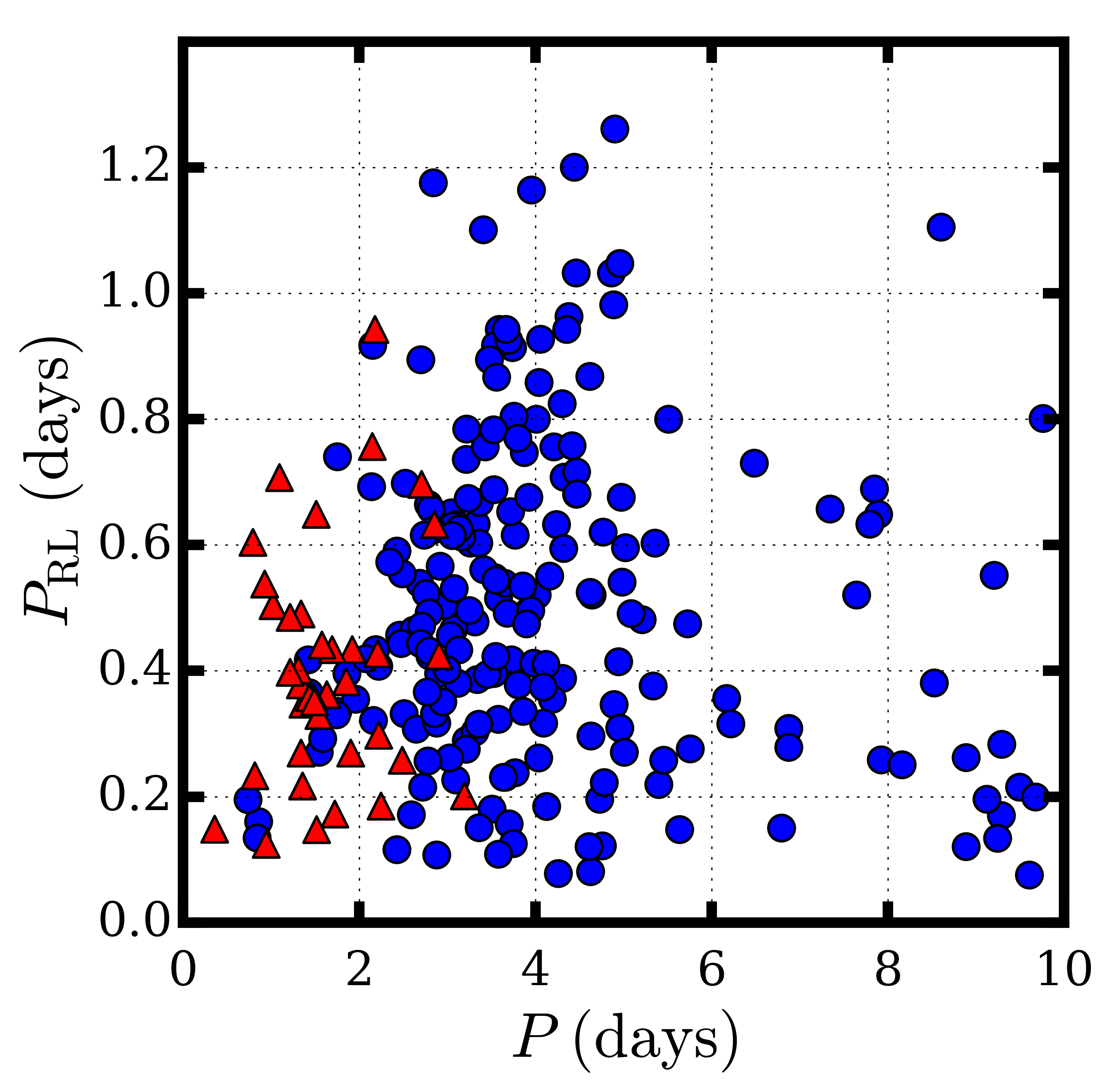}
\caption{Roche limit periods $P_{\rm RL}$ (Equation \ref{eq:P_RL}) vs. orbital periods $P$ for many short-period planets. Red triangles indicate planets for which tidal interactions with the host star may bring a planet with zero orbital eccentricity into its Roche limit in less than 1 Gyr (assuming a modified tidal dissipation parameter for the star $Q_{\rm s}^\prime = 10^7$ -  \citealp{2012ApJ...751...96P}) -- see ``Tides in star-planet systems'' chapter.}
\label{fig:Jackson-Carlberg_fig3}
\end{figure}

For planets (and stars) composed of slowly moving fluid, contours of $\Phi$ coincide closely with density contours, and so the surface of a planet just filling its Roche lobe will correspond very nearly to the Roche lobe. To decide whether a planet is losing mass, we can compare the planet's mean radius $R_{\rm p}$ to $R_{\rm RL}$. Incorporating \hbindex{Kepler's Third Law}, \citet{2013ApJ...773L..15R} recast Equation \ref{eq:small_q_roche_lobe} to estimate the orbital period for the Roche limit, $P_{\rm RL}$, from the corresponding semi-major axis, $a_{\rm RL}$, for a planet with a bulk density $\rho_{\rm p}$: 
\begin{equation}
\label{eq:P_RL}
P_{\rm RL} \approx \sqrt{\frac{3\ \pi}{\left( 0.49 \right)^{3} G \rho_{\rm p}}} \approx 9.6\ {\rm hrs} \left( \frac{\rho_{\rm p}}{1\ {\rm g\ cm^{-3}}} \right)^{-1/2}. 
\end{equation}
Figure \ref{fig:Jackson-Carlberg_fig3} compares the current periods $P$ to $P_{\rm RL}$ for several short-period planets.

Importantly, the classical Roche limit here involves a number of assumptions, including treating the planet and star as point masses, assuming a circular orbit \citep{2008ApJ...678.1396J} and synchronized planetary rotation \citep{2002A&A...385..166S}, neglecting the effects of additional bodies \citep{2012CeMDA.113..215V, 2015MNRAS.450.4505H}, and neglecting tensile strength and friction \citep{1998Icar..134...47R, 1999Icar..142..525D}.

\section{How are Planetary Bodies Disrupted?}
\label{sec:How_Disrupted}
Once some portion of the planet's surface or atmosphere contacts the Roche lobe, it will begin losing mass (\hbindex{Roche-lobe overflow} or RLO).  Whether RLO proceeds stably or not depends on how the planet and its orbit respond to the mass loss. For dense planets with $a_{\rm RL}$ below the stellar photosphere, other processes determine how the planet is disrupted.

	\subsection{Roche-Lobe Overflow}
For short-period exoplanets, \hbindex{tidal interactions} with the host star usually transfer orbital angular momentum to the star's rotation and reduce the semi-major axis. Even for zero eccentricity, tides can bring  hot Jupiters ($P\sim$ days) into Roche-lobe contact within Gyrs \citep{2008ApJ...678.1396J}. Smaller planets raise smaller tides, driving slower orbital decay, but Earth-sized, rocky planets have been found with $P \sim$ hrs and may also be accreted in Gyrs \citep{2013ApJ...774...54S, 2013ApJ...779..165J, 2014ApJ...787...47S, 2016AJ....152...47A}.

Material leaks out primarily through the L1 \hbindex{Lagrange point} between the planet and the star. As the mass escapes, conservation of angular momentum can form a thin \hbindex{accretion disk} around the star. Viscous stresses within the disk and tidal interactions with the disrupting planet can transfer some of the disk's angular momentum back to the planet, driving the material inward \citep{Ritter1988Turning}. 

Torques between the planet and disk act in the opposite direction as the torque from the stellar tide \citep{1979MNRAS.186..799L}, but mass loss would stop if the planet stopped filling its Roche lobe. Consequently, the disk torque cannot drive the planet beyond the Roche limit. However, if the tide raised on the host star pulled the planet inward through the Roche limit, the mass loss would increase \citep{Ritter1988Turning}, increasing the mass in the accretion disk and the strength of the disk's torque. 
The resulting balance, stable RLO \citep{Priedhorsky1988Tidal}, keeps the planet's radius $R_{\rm p} \approx R_{\rm RL}$. If the planet's density drops as it loses mass, $P_{\rm RL}$ increases (Equation \ref{eq:P_RL}), and the gas disk would drive the planet out with the Roche limit \citep{2015ApJ...813..101V,2016CeMDA.126..227J}.

This balance relates the planet's orbital evolution to its mass and radius evolution:
\begin{equation}
\left( \frac{\dot{M}_{\rm p}}{M_{\rm p}} \right) = -\frac{1}{2}
\eta^{-1} \left[ \frac{1}{a}\left(\frac{\partial a}{\partial
t}\right)_{\rm star} + \left( \frac{\partial \ln R_{\rm p}}{\partial
t} \right)_{\rm M_{\rm p}} \right],
\label{eq:RLO_mass_loss_rate}
\end{equation}
\begin{equation}
\left( \frac{\dot{a}}{a} \right) = \eta^{-1} \left[\left( \frac{1}{3} - \xi \right) \frac{1}{a}\left(\frac{\partial a}{\partial t}\right)_{\rm star} + \left( 1 - \delta \gamma \right) \left( \frac{\partial \ln R_{\rm p}}{\partial t} \right)_{\rm M_{\rm p}} \right],
\label{eq:a_evolution_rate}
\end{equation}
where $M_{\rm p} \ll M_{\rm s}$. 

Here, $\xi = \partial \ln R_{\rm p}/\partial \ln M_{\rm p}$. The escaping gas has a specific angular momentum $\gamma \sqrt{G M_{\rm s} a}$, but the angular momentum is not necessarily returned to the planet. Gas may escape from the system, driven by stellar wind \citep{1995A&A...297..727B}, or angular momentum is transferred to the star during accretion. The parameter $\delta$ expresses the fraction of gas that does not return its angular momentum: $\delta = 0$ means all is returned, while $\delta = 1$ means none is. These parameters combine to give $\eta = \xi/2 - \delta \gamma + 5/6$ ($5/6$ comes from the Roche lobe's mass dependence). 

A vast literature exists that provides many formulations for tidal interactions in binary systems (see \citealt{2008CeMDA.101..171F} for a thorough review). One well-worn model \citep{1966Icar....5..375G} gives
\begin{equation}
\label{eq:dadt_star}
\left(\frac{\partial a}{\partial t}\right)_{\rm star} = -\frac{9}{2} \left( \frac{G}{M_{\rm s}} \right)^{1/2} \frac{R_{\rm s}^5 M_{\rm p}}{Q_{\rm s}^\prime} a^{-11/2},
\end{equation}
where $Q_{\rm s}^\prime$ is the modified \hbindex{tidal dissipation parameter} \citep{2008CeMDA.101..171F}. Unfortunately, the details of tidal dissipation for stars and gaseous planets are not well understood, and predictions for $Q_{\rm s}^\prime$ span a wide range \citep[cf.][]{2014ARA&A..52..171O}. Consequently, the timescales for tidal decay and RLO are poorly known, but the outcomes for stable RLO should be  insensitive to the rates.

With a \hbindex{mass-radius relation}, Equations \ref{eq:RLO_mass_loss_rate} and \ref{eq:a_evolution_rate} can be integrated to model the evolution of a disrupting gaseous planet.  \citet{2015ApJ...813..101V} and \citet{2016CeMDA.126..227J} employed the Modules in Stellar Astrophysics (\hbindex{MESA}) suite \citep{2011ApJS..192....3P, 2013ApJS..208....4P, 2015ApJS..220...15P} to model disrupting hot Jupiters with a variety of initial conditions, and Figure \ref{fig:Jackson-Carlberg_fig4} illustrates the results. For a hot Jupiter $\xi \approx 0$ \citep{Fortney2007Planetary}, so mass loss initially reduces $\rho_{\rm p}$, increases $P_{\rm RL}$, and drives planets out. Eventually, mass loss removes most of the gaseous envelope, $\rho_{\rm p}$ and $P_{\rm RL}$ drop, and the tide can draw the planet back in. The outcome depends on the mass of the planet's solid core, not on initial conditions (assuming they allow for RLO): the smaller the core, the more orbital expansion.

\begin{figure}
\caption{Figure 4 from \citet{2016CeMDA.126..227J}. Not included due to copyright restrictions.}
\label{fig:Jackson-Carlberg_fig4}
\end{figure}

\citet{2016CeMDA.126..227J} found that the gaseous planets that reach their Roche limits can completely shed their atmospheres, resulting in significant orbital expansion (which slows disruption and may save the remnant from accretion) or rapid accretion of the planet by the star. 

Stable RLO requires a specific relationship between evolution of $R_{\rm p}$ and of $R_{\rm RL}$:
\begin{equation}
\delta \gamma < \xi/2 + 5/6.
\label{eq:stability_condition}
\end{equation}
If this inequality is not satisfied, unstable RLO may result. For example, if $\gamma \approx 1$ and $\xi \approx 0$, stability requires no more than 5/6 (= $0.8\bar{3}$) of the mass escaping from the planet be lost without returning its angular momentum. Loss of orbital angular momentum is inevitable -- \citet{2012MNRAS.425.2778M} pointed out that material accreted at the stellar surface carries non-negligible angular momentum which will not be returned to the orbit, translating into $\delta \gamma = \sqrt{R_\star/a_{\rm RL}}$. For Jupiter orbiting the Sun, $a = a_{\rm Roche} \approx 0.01\ {\rm AU} \approx 2\ R_{\rm Sun}$, and $\delta \gamma = \sqrt{1/2}$. 

\citet{2015ApJ...813..101V} explored loss of escaped material and found instability was likely, while \citet{2017MNRAS.465..149J} found rocky planets between 1 and 10 Earth masses $M_{\rm Earth}$ are likely to undergo unstable mass transfer. During unstable RLO, the mass loss timescale is comparable to the orbital period, but what happens next is unclear. \citet{2015ApJ...813..101V} speculated that the escaped material may form a torus orbiting the host star, and interactions with the denuded planet may eject the torus, driving the planet into its Roche limit or the star. 
 
    \subsection{Disruption Below the Stellar Photosphere}
The photosphere for a Sun-like, main sequence star corresponds to the Roche limit for $\rho_{\rm p} = 8$~${\rm g~cm^{-3}}$, denser than nearly all super-Earths found by the \hbindex{Kepler mission} \citep{2014ApJ...783L...6W}. Planets orbiting lower density, red giant stars, though, may enter the stellar photosphere relatively unscathed - Figure \ref{fig:Jackson-Carlberg_fig1}. 

Several outcomes are possible for such planets. The most massive companions may unbind the stellar envelopes, but only for \hbindex{asymptotic giant branch stars}, when stars reach their largest radii and thus lowest binding energy \citep{2006MNRAS.370.2004N}. In smaller red giants, companions can (1) enter into a common envelope stage if $M_{\rm p} \geq~$5--20$~M_{\rm Jup}$, with the final mass of the accreting companion dependent on the star's envelope mass; (2) fully evaporate in the stellar envelope; or (3) impact the stellar core \citep{1984MNRAS.208..763L} .

For the latter two cases, evaporation within the star will occur near the depth where the temperature of the stellar interior exceeds the \hbindex{virial temperature} of the planet  $T_{\rm v}$ \citep{1999MNRAS.304..925S}:
\begin{equation}
\label{eq:virial}
T_{\rm v} = \frac{GM_{\rm p}\mu_{\rm p} m_{\rm H}}{kR_{\rm p}},
\end{equation}
where $\mu_{\rm p}$ is the planet's mean molecular weight, $m_{\rm H}$ hydrogen's mass, and $k$ Boltzmann's constant. \cite{2016ApJ...829..127A} modeled dissolution for a range of  companion planets and brown dwarfs and stellar masses and metallicities, and found that companions with $M_{\rm p} \leq 15~M_{\rm Jup}$ will evaporate in the stellar envelope but more massive companions would probably impact the stellar core.


\section{When are Planetary Bodies Disrupted?}
\label{sec:When_Disrupted}
Planetary accretion may occur from the pre-main sequence through the white dwarf stage, and different physical processes drive the accretion in different stages. The composition and rotation of a host star and orbital architecture of a planetary system may be long-lived signatures of planetary accretion, although unequivocal evidence is lacking for many stages. 

	\subsection{During the Pre- and Main Sequence}

In principle, accretion of planetary material changes a star's composition, an idea originally invoked to explain the \hbindex{planet-metallicity correlation} \citep{2005ApJ...622.1102F}. The stellar mass fraction $X_i$ for an element $i$ is given as \cite{1999MNRAS.308.1133S}
\begin{equation}
\label{eq:X_i}
X_i = \frac{X_i^{\rm CZ}M_{\rm CZ}+X_i^{\rm acc}M_{\rm acc}}{M_{\rm acc}+M_{\rm CZ}},
\end{equation}
where `CZ' refers to the \hbindex{stellar convection zone} and `acc' to the accreted material, and $M$ is the associated total mass. 

The changes to the photospheric abundances depend on the amount of accreted material and how much its composition differs from the star's. For hydrogen-dominated, giant planets, the least stellar-like compositions are found in their cores. The convection zone in Sun-like stars makes up a small fraction of the stars' masses, and so a little material can cause large local abundance enhancements. The fact that the planet-metallicity correlation does not depend on convective envelope mass strongly suggests the correlation does not arise from planetary accretion  \citep{2005ApJ...622.1102F}.

However, accretion during the pre- and main sequence may impart other chemical signatures. Recent work suggests the last gas accreted onto stars is depleted in planet-forming materials, which are probably locked-up in terrestrial planets or gas giant's cores. \cite{melendez09} and \cite{2009A&A...508L..17R} found that the absence of $\sim 4~M_\oplus$ of rocky material could explain a depletion in refractory elements in the Sun's photosphere compared to solar twins, fueling research into abundance comparisons between binary stars for which only one hosts a planet. For example, the 16 Cyg binary system is composed of two solar analogs, and one hosts a planet \citep{1997ApJ...483..457C}. The co-evality and nearly identical masses of the stars makes 16 Cyg ideal for measuring abundance differences, and \cite{2014ApJ...790L..25T} found a telling deficiency in the planet-hosting star corresponding to 1.5--6~$M_\oplus$ of missing metals. 

Planetary accretion may induce downward mixing \citep{2012ApJ...744..123T}, effectively erasing expected abundance enhancements (Equation \ref{eq:X_i}) but propitiously, this mixing can also create a new signature. The convection zone is the only region in Sun-like stars where fragile elements, like \hbindex{lithium}, can survive, and downward mixing  carries those elements to hotter regions, where they are destroyed, permanently depleting their surface abundances. \cite{2015A&A...584A.105D} argued that this scenario could explain the discrepant lithium abundances in the 16 Cyg system if the planet-hosting star, which has less lithium, accreted less than an Earth's mass of material.

Some migration likely brings short-period planets in from where they form, and two types have been considered: \hbindex{gas disk migration} and \hbindex{dynamical excitation}, followed by \hbindex{tidal circularization}. In later stages, planets may enjoy long-lived stability, but, for short-period planets, tidal interactions can drive significant orbital decay  -- Figure \ref{fig:Jackson-Carlberg_fig3}. Additional evidence for planetary accretion may come from considering these dynamical processes.

		\subsubsection{Due to Gas Disk Migration}
\hbindex{Gas disk migration} involves gravitational interactions between a planet and its disk and can drive significant orbital evolution during the disk's lifetime, $\leq$ 10 Myrs \citep{2013MNRAS.434..806B, 1998ApJ...500..428T, 2009AREPS..37..321C}. \citet{1996Natur.380..606L} suggested a young star's magnetosphere might clear the protoplanetary disk several stellar radii from the star, allowing gas disk migration to deposit planets in orbits near the disk's inner edge \citep{1979MNRAS.186..799L, 1979ApJ...233..857G}. Coupling between the  magnetosphere and disk may synchronize the star's rotation to the innermost disk \citep{1991ApJ...370L..39K}; thus, a young star's rotation period may indicate the orbital period of the disk's inner edge and the stopping point for gas disk migration. \citet{2009IAUS..258..363I} found young stars' rotation periods  range from $<$ 1 day up to $\sim$ 10 days, suggesting that disk migration may bring planets very close in. Indeed, \citet{2012ApJ...755...42V} discovered a possible hot Jupiter in a 10-hour orbit around a T Tauri star, likely in RLO \citep{2017ApJ...835..145J}. 

		\subsubsection{Due to Dynamical Excitation}
Under an alternative scenario, short-period planets formed in roughly circular orbits with initial semi-major axes $\geq$ 1 AU. Interactions with other planets or companion stars via Kozai resonances \citep{2007ApJ...671..872Z, 2016ARA&A..54..441N} or dynamical instabilities \citep{1996Sci...274..954R, 2008ApJ...686..580C} excited the planets' \hbindex{orbital inclinations} and eccentricities to large values, giving small enough pericenters for tidal circularization to shrink the orbits. Given the potentially chaotic nature of these interactions, production of short-period planets is only one outcome \citep{2008ApJ...686L..29M, 2008ApJ...678..498N}. Often, a planet is scattered into a Roche-limit crossing orbit, resulting in disruption and accretion \citep{2011ApJ...732...74G, 2013ApJ...762...37L}.  Since tidal circularization is dominated by dissipation within the planet \citep{2008ApJ...678.1396J}, it can also heat and inflate a gaseous planet to its Roche limit \citep{2003ApJ...588..509G}.

Since dynamical excitation can significantly misalign the stellar rotation and orbital angular momentum vectors, the \hbindex{spin-orbit alignment} for planetary systems may reveal their histories. For many systems, \citet{2012ApJ...757...18A} compiled prior and new estimates of stellar obliquity angles as projected on the plane of the sky (cf. \citealp{2009ApJ...696.1230F}) and found the angles were more broadly distributed for hotter stars than for cooler stars. Since tidal interactions may be more effective at reducing obliquities for cooler than hotter stars, this distribution may point to the dynamical excitation mode of short-period planet production \citep{2010ApJ...718L.145W}. Other interpretations of the obliquity distribution do not require the dynamical excitation (and concomitant accretion) \citep{2012Natur.491..418B, 2013Sci...342..331H}.

Other aspects of the orbital architectures of planetary systems may also suggest planetary accretion via dynamical excitation. \cite{2007ApJ...660..823M} demonstrated that material around a soon-to-be hot Jupiter migrating through a gas disk can coalesce into low-mass companions, but \citet{2015ApJ...808...14M} pointed out that dynamical excitation of the planet would instead empty the inner system of low-mass planets by driving many into the host star. The apparent lack of sibling planets in hot Jupiter systems supports  this latter scenario \citep{2012PNAS..109.7982S}, although the idea that hot Jupiters tend to be ``lonely'' has been challenged \citep{2016ApJ...825...62S}.

		\subsubsection{Due to Tidal Decay}
After the disk has dissipated, tidal interactions with a short-period planet and loss of angular momentum via \hbindex{stellar wind} dominate a star's rotational evolution: the former tends to synchronize the rotation with the planet's orbital motion, usually increasing the spin rate  \citep{2002ApJ...573..829M, 2008ApJ...678.1396J, 2011MNRAS.415..605B}, while the latter slows the spin \citep{1972ApJ...171..565S, 1987ApJ...318..337S, 1988ApJ...333..236K, 1996ApJ...462..746B}. Drawing on \citet{1973ApJ...180..307C}, \citet{2009ApJ...692L...9L} and \citet{2015MNRAS.446.3676A} showed the total angular momentum (rotational + orbital) for many exoplanetary systems lies below a threshold value for stability, meaning those planets are doomed to RLO or accretion - Figure \ref{fig:Jackson-Carlberg_fig5}. Since this result holds whatever the rate of tidal decay, determining whether decay occurs on Gyr timescales requires additional observations. 

\begin{figure}
\caption{Figure 3 from \citet{2015MNRAS.446.3676A}. Not included due to copyright restrictions.}
\label{fig:Jackson-Carlberg_fig5}
\end{figure}

\citet{2006ApJ...638L..45F} pointed out that the dynamical excitation production mode should give an initial cut-off in orbital semi-major axes at twice the Roche limit, and the distribution does show such a cut-off, albeit with some of the shortest-period planets interior to this cut-off. \citet{2014ApJ...787L...9V} interpreted this distribution as evidence for tidal decay subsequent to dynamical excitation and tidal circularization. 

Evidence of tidal spin-up and planetary engulfment also appears in the distribution of planetary orbits for stars of different rotation periods. \cite{2013ApJ...775L..11M} measured rotation periods of main sequence planet hosts observed by the \hbindex{Kepler mission} and found that the faster the stars rotated, the farther out the innermost planets were found. \cite{2014ApJ...786..139T} interpreted this relationship as evidence that the close-in planets of those fast-rotating stars had been engulfed.

	\subsection{During the Post-Main Sequence}
    
During the \hbindex{red giant phase}, expansion of the stellar envelope enhances tidal decay of planetary orbits (Equation \ref{eq:dadt_star}), which competes with orbital expansion from stellar mass loss. Figure \ref{fig:Jackson-Carlberg_fig1} shows that many known exoplanets may eventually be engulfed, and numerous studies have explored the survival limits around red giants (e.g., \citealt{2007ApJ...661.1192V,2009ApJ...705L..81V,2009ApJ...700..832C, 2011ApJ...737...66K}). A corollary of this observation is that many  current red giant stars have engulfed one or more planetary companions. 

\cite{1999MNRAS.308.1133S} was one of the earliest comprehensive calculations of the signatures of planet engulfment by an evolved solar-type star. They modeled engulfment as an accretion rate deep in the star and predicted enhanced rotation rates, changes to the composition of the stellar atmosphere, reemergence of a dynamo with associated X-ray emission, and ejection of shellular material with associated infrared emission. 

The rotation rates of red giants are generally slow and unmeasurable in many cases. Enhanced rotation is thus a telltale signature of engulfment, provided that the stellar mass is known to be low (since low-mass stars are slow main sequence rotators) and that interactions with stellar companions can be ruled out. \cite{2009ApJ...700..832C} demonstrated that known exoplanet systems have sufficiently large angular momentum that planetary accretion could create rapid stellar  rotation on the red giant branch. However, this fast rotation signature will not persist throughout the red giant phase because of the steadily increasing stellar radius. Furthermore, its detection requires a favorable orientation of the stellar rotation axis to the observer's line of sight. 

Surface composition changes are expected to be modest for engulfment by evolved stars because of their large convection zones (large $M_{\rm CZ}$ in Equation \ref{eq:X_i}). A notable exception is the group of fragile elements, Li, Be, and B, which are destroyed by proton-capture reactions throughout the star's life and should be depleted in the atmospheres of red giants that have not engulfed a planet. Lithium is the element most affected by this process. For stars undergoing simple \hbindex{first-dredge up mixing}, the depletion factor is approximately a few tens but can rise to 100,000 for deep circulation between the stellar envelope and the interior \citep{1999ApJ...510..217S}. In such stars, the amount of Li deposited in the stellar envelope by an engulfed planet could greatly exceed the Li already present.
 
\cite{1967Obs....87..238A} first predicted that engulfed planets could increase the surface Li in red giants, and Li-enriched red giants were subsequently discovered \citep{1982ApJ...255..577W}.  However, some of these red giants have Li exceeding expectations from engulfment. Paradoxically, stars with deep circulation may be able  to replenish the surface lithium. Stars synthesize Li from $^3$He at depths where the Li is almost instantly destroyed, but fast circulation of material between these depths and the surface can conceivably salvage  newly created Li  \citep{1999ApJ...510..217S}. Thus, the ability to use Li as an engulfment signature is limited by how well a given red giant star's {\it expected} Li abundance can be constrained.

Although identifying individual planet-engulfing red giants remains elusive, evidence of engulfment appears in ensemble studies. \cite{2002AJ....123.2703D} noted that fast-rotating red giants were much more likely to be Li-rich than slow red giants, and \cite{2012ApJ...757..109C} demonstrated a global enhancement of Li in fast-rotators, consistent with simultaneous increases in both rotation and Li from engulfment. 

Further evidence for engulfment may be seen in the distribution of planetary companions to evolved stars: there is a deficiency in short-period companions (with semi-major axes between 0.06 to 0.5 AU) compared to main sequence populations. However, the extent of the deficiency may not be consistent with tidal engulfment since many red giants have not grown large enough to engulf planets out to 0.5 AU. A few recent discoveries, notably HIP13044 at 0.116~AU \citep{2010Sci...330.1642S} and Kepler-91 at 0.07~AU \citep{2014A&A...562A.109L}, include planets passing within a few stellar radii at pericenter passage. In any case, the protracted desert of planets between 0.06 and 0.5 AU around evolved stars remains unexplained. 

	\subsection{In the Stellar Graveyard}
Not even stellar death can stop the accretion of planetary material. Roughly the size of Earth but with masses comparable to the Sun, \hbindex{white dwarfs} are extremely dense ($\geq 10^4$ g/cm$^3$) and hot ($\sim$ 10,000 K) objects primarily composed of carbon and oxygen covered with thin, stratified atmospheres of helium and hydrogen. Given the large densities of white dwarfs, disruption of planetary bodies likely results in the formation of an accretion disk, with the disk materials transitioning from solid particles in the outermost regions to vapor in the innermost regions nearest the white dwarf \citep{2014MNRAS.445.2244V}. 

The observational evidence supporting accretion of planetary material onto white dwarfs is more robust than for any other stage of stellar evolution. Elements heavier than He should settle out of the atmosphere in just weeks or less due to the white dwarf's strong gravitational field, except in white dwarfs hotter than $\sim 20,000$K \citep{1995ApJS...99..189C}, where \hbindex{radiative levitation} counteracts the settling. Nevertheless, absorption features from metals have been detected in white dwarfs cooler than 20,000 K, and these metals must have been recently or are continually being accreted. The sources of these metals are thought to be disrupted asteroids and/or comets \citep{2007ApJ...671..872Z}, an idea supported by the infrared excesses seen around white dwarfs. The outer radii of the disks inferred from the infrared are $\sim R_\odot$, or approximately the Roche limit for asteroidal bodies \citep{2003ApJ...584L..91J}. How these asteroids and comets are injected into Roche-lobe crossing orbits remains unclear, but the shifting of resonant interactions between a distant asteroid or Kuiper belt and a perturbing planet as the result of stellar mass loss may be responsible \citep{2011MNRAS.414..930B, 2012ApJ...747..148D}.

The measured relative abundances provide estimates of the bulk composition of the accreted material, and the resulting elemental ratios can be compared to classes of objects in the solar system to infer the properties of the accreted object. For example, relative surpluses or deficiencies in Fe to Si have been interpreted as accreted cores or crusts of a differentiated body. There is even evidence for the accretion of water-rich planetary debris, discovered by accounting for the various mineralogical carriers of oxygen (e.g., SiO$_2$) and attributing the observed excess O to H$_2$O \citep{2013Sci...342..218F}. 

\begin{acknowledgement}
This work has made use of the Exoplanet Orbit Database and the Exoplanet Data Explorer at exoplanets.org.
\end{acknowledgement}

\bibliographystyle{spbasicHBexo}  
\bibliography{Jackson-Carlberg} 

\begin{thebibliography}{103}
\providecommand{\natexlab}[1]{#1}
\providecommand{\url}[1]{{#1}}
\providecommand{\urlprefix}{URL }
\expandafter\ifx\csname urlstyle\endcsname\relax
  \providecommand{\doi}[1]{DOI~\discretionary{}{}{}#1}\else
  \providecommand{\doi}{DOI~\discretionary{}{}{}\begingroup
  \urlstyle{rm}\Url}\fi
\providecommand{\eprint}[2][]{\url{#2}}

\bibitem[{{Adams} et~al.(2016){Adams}, {Jackson}, and
  {Endl}}]{2016AJ....152...47A}
{Adams} ER, {Jackson} B {Endl} M (2016) {Ultra-short-period Planets in K2
  SuPerPiG Results for Campaigns 0-5}. \aj 152:47

\bibitem[{{Adams} and {Bloch}(2015)}]{2015MNRAS.446.3676A}
{Adams} FC {Bloch} AM (2015) On the stability of extrasolar planetary systems
  and other closely orbiting pairs. Monthly Notices of the Royal Astronomical
  Society 446:3676--3686,
  \urlprefix\url{http://dx.doi.org/10.1093/mnras/stu2397}

\bibitem[{Aguilera-G{\'o}mez et~al.(2016)Aguilera-G{\'o}mez, Chanam{\'e},
  Pinsonneault, and Carlberg}]{2016ApJ...829..127A}
Aguilera-G{\'o}mez C, Chanam{\'e} J, Pinsonneault MH Carlberg JK (2016) {On
  Lithium-rich Red Giants. I. Engulfment of Substellar Companions}. The
  Astrophysical Journal 829(2):127

\bibitem[{{Albrecht} et~al.(2012){Albrecht}, {Winn}, {Johnson}, {Howard},
  {Marcy}, {Butler}, {Arriagada}, {Crane}, {Shectman}, {Thompson}, {Hirano},
  {Bakos}, and {Hartman}}]{2012ApJ...757...18A}
{Albrecht} S, {Winn} JN, {Johnson} JA et~al. (2012) {Obliquities of Hot Jupiter
  Host Stars: Evidence for Tidal Interactions and Primordial Misalignments}.
  \apj 757:18

\bibitem[{Alexander(1967)}]{1967Obs....87..238A}
Alexander JB (1967) {A possible source of lithium in the atmospheres of some
  red giants}. The Observatory 87:238--240

\bibitem[{{Barnes} and {Sofia}(1996)}]{1996ApJ...462..746B}
{Barnes} S {Sofia} S (1996) {On the Origin of the Ultrafast Rotators in Young
  Star Clusters}. \apj 462:746

\bibitem[{{Batygin}(2012)}]{2012Natur.491..418B}
{Batygin} K (2012) {A primordial origin for misalignments between stellar spin
  axes and planetary orbits}. \nat 491:418--420

\bibitem[{{Bell} et~al.(2013){Bell}, {Naylor}, {Mayne}, {Jeffries}, and
  {Littlefair}}]{2013MNRAS.434..806B}
{Bell} CPM, {Naylor} T, {Mayne} NJ, {Jeffries} RD {Littlefair} SP (2013)
  {Pre-main-sequence isochrones - II. Revising star and planet formation
  time-scales}. Monthly Notices of the Royal Astronomical Society 434:806--831,
  \urlprefix\url{http://dx.doi.org/10.1093/mnras/stt1075}

\bibitem[{{Bloecker}(1995)}]{1995A&A...297..727B}
{Bloecker} T (1995) {Stellar evolution of low and intermediate-mass stars. I.
  Mass loss on the AGB and its consequences for stellar evolution.} \aap
  297:727

\bibitem[{{Bonsor} et~al.(2011){Bonsor}, {Mustill}, and
  {Wyatt}}]{2011MNRAS.414..930B}
{Bonsor} A, {Mustill} AJ {Wyatt} MC (2011) {Dynamical effects of stellar
  mass-loss on a Kuiper-like belt}. \mnras 414:930--939

\bibitem[{{Brown} et~al.(2011){Brown}, {Collier Cameron}, {Hall}, {Hebb}, and
  {Smalley}}]{2011MNRAS.415..605B}
{Brown} DJA, {Collier Cameron} A, {Hall} C, {Hebb} L {Smalley} B (2011) {Are
  falling planets spinning up their host stars?} \mnras 415:605--618

\bibitem[{Carlberg et~al.(2009)Carlberg, Majewski, and
  Arras}]{2009ApJ...700..832C}
Carlberg JK, Majewski SR Arras P (2009) {The Role of Planet Accretion in
  Creating the Next Generation of Red Giant Rapid Rotators}. The Astrophysical
  Journal 700(1):832--843

\bibitem[{Carlberg et~al.(2012)Carlberg, Cunha, Smith, and
  Majewski}]{2012ApJ...757..109C}
Carlberg JK, Cunha K, Smith VV Majewski SR (2012) {Observable Signatures of
  Planet Accretion in Red Giant Stars. I. Rapid Rotation and Light Element
  Replenishment}. The Astrophysical Journal 757(2):109

\bibitem[{{Chambers}(2009)}]{2009AREPS..37..321C}
{Chambers} JE (2009) {Planetary Migration: What Does It Mean for Planet
  Formation?} Annual Review of Earth and Planetary Sciences 37:321--344

\bibitem[{{Chatterjee} et~al.(2008){Chatterjee}, {Ford}, {Matsumura}, and
  {Rasio}}]{2008ApJ...686..580C}
{Chatterjee} S, {Ford} EB, {Matsumura} S {Rasio} FA (2008) {Dynamical Outcomes
  of Planet-Planet Scattering}. \apj 686:580-602

\bibitem[{{Chayer} et~al.(1995){Chayer}, {Fontaine}, and
  {Wesemael}}]{1995ApJS...99..189C}
{Chayer} P, {Fontaine} G {Wesemael} F (1995) {Radiative Levitation in Hot White
  Dwarfs: Equilibrium Theory}. \apjs 99:189

\bibitem[{{Cochran} et~al.(1997){Cochran}, {Hatzes}, {Butler}, and
  {Marcy}}]{1997ApJ...483..457C}
{Cochran} WD, {Hatzes} AP, {Butler} RP {Marcy} GW (1997) {The Discovery of a
  Planetary Companion to 16 Cygni B}. \apj 483:457--463

\bibitem[{{Counselman}(1973)}]{1973ApJ...180..307C}
{Counselman} CC III (1973) {Outcomes of Tidal Evolution}. \apj 180:307--316

\bibitem[{{Davidsson}(1999)}]{1999Icar..142..525D}
{Davidsson} BJR (1999) {Tidal Splitting and Rotational Breakup of Solid
  Spheres}. \icarus 142:525--535

\bibitem[{Deal et~al.(2015)Deal, Richard, and Vauclair}]{2015A&A...584A.105D}
Deal M, Richard O Vauclair S (2015) {Accretion of planetary matter and the
  lithium problem in the 16 Cygni stellar system}. A{\&}A 584:A105

\bibitem[{{Debes} et~al.(2012){Debes}, {Walsh}, and
  {Stark}}]{2012ApJ...747..148D}
{Debes} JH, {Walsh} KJ {Stark} C (2012) {The Link between Planetary Systems,
  Dusty White Dwarfs, and Metal-polluted White Dwarfs}. \apj 747:148

\bibitem[{Drake et~al.(2002)Drake, de~la Reza, da~Silva, and
  Lambert}]{2002AJ....123.2703D}
Drake NA, de~la Reza R, da~Silva L Lambert DL (2002) {Rapidly Rotating
  Lithium-rich K Giants: The New Case of the Giant PDS 365}. The Astronomical
  Journal 123(5):2703--2714

\bibitem[{{Eggleton}(1983)}]{1983ApJ...268..368E}
{Eggleton} PP (1983) {Approximations to the radii of Roche lobes}. \apj 268:368

\bibitem[{{Fabrycky} and {Winn}(2009)}]{2009ApJ...696.1230F}
{Fabrycky} DC {Winn} JN (2009) {Exoplanetary Spin-Orbit Alignment: Results from
  the Ensemble of Rossiter-McLaughlin Observations}. \apj 696:1230--1240

\bibitem[{{Farihi} et~al.(2013){Farihi}, {G{\"a}nsicke}, and
  {Koester}}]{2013Sci...342..218F}
{Farihi} J, {G{\"a}nsicke} BT {Koester} D (2013) {Evidence for Water in the
  Rocky Debris of a Disrupted Extrasolar Minor Planet}. Science 342:218--220

\bibitem[{{Ferraz-Mello} et~al.(2008){Ferraz-Mello}, {Rodr{\'{\i}}guez}, and
  {Hussmann}}]{2008CeMDA.101..171F}
{Ferraz-Mello} S, {Rodr{\'{\i}}guez} A {Hussmann} H (2008) {Tidal friction in
  close-in satellites and exoplanets: The Darwin theory re-visited}. Celestial
  Mechanics and Dynamical Astronomy 101:171--201

\bibitem[{{Fischer} and {Valenti}(2005)}]{2005ApJ...622.1102F}
{Fischer} DA {Valenti} J (2005) {The Planet-Metallicity Correlation}. \apj
  622:1102--1117

\bibitem[{{Ford} and {Rasio}(2006)}]{2006ApJ...638L..45F}
{Ford} EB {Rasio} FA (2006) {On the Relation between Hot Jupiters and the Roche
  Limit}. \apjl 638:L45--L48

\bibitem[{{Fortney} et~al.(2007){Fortney}, {Marley}, and
  {Barnes}}]{Fortney2007Planetary}
{Fortney} JJ, {Marley} MS {Barnes} JW (2007) Planetary radii across five orders
  of magnitude in mass and stellar insolation: Application to transits.
  Astrophys J 659(2):1661--1672,
  \urlprefix\url{http://dx.doi.org/10.1086/512120}

\bibitem[{{Goldreich} and {Soter}(1966)}]{1966Icar....5..375G}
{Goldreich} P {Soter} S (1966) {Q in the Solar System}. \icarus 5:375--389

\bibitem[{{Goldreich} and {Tremaine}(1979)}]{1979ApJ...233..857G}
{Goldreich} P {Tremaine} S (1979) {The excitation of density waves at the
  Lindblad and corotation resonances by an external potential}. \apj
  233:857--871

\bibitem[{{Gu} et~al.(2003){Gu}, {Lin}, and
  {Bodenheimer}}]{2003ApJ...588..509G}
{Gu} PG, {Lin} DNC {Bodenheimer} PH (2003) {The Effect of Tidal Inflation
  Instability on the Mass and Dynamical Evolution of Extrasolar Planets with
  Ultrashort Periods}. \apj 588:509--534

\bibitem[{{Guillochon} et~al.(2011){Guillochon}, {Ramirez-Ruiz}, and
  {Lin}}]{2011ApJ...732...74G}
{Guillochon} J, {Ramirez-Ruiz} E {Lin} D (2011) {Consequences of the Ejection
  and Disruption of Giant Planets}. \apj 732:74

\bibitem[{{Han} et~al.(2014){Han}, {Wang}, {Wright}, {Feng}, {Zhao},
  {Fakhouri}, {Brown}, and {Hancock}}]{2014PASP..126..827H}
{Han} E, {Wang} SX, {Wright} JT et~al. (2014) {Exoplanet Orbit Database. II.
  Updates to Exoplanets.org}. \pasp 126:827

\bibitem[{{Hansen} and {Zink}(2015)}]{2015MNRAS.450.4505H}
{Hansen} BMS {Zink} J (2015) {On the potentially dramatic history of the
  super-Earth {$\rho$} 55 Cancri e}. \mnras 450:4505--4520

\bibitem[{{Huber} et~al.(2013){Huber}, {Carter}, {Barbieri}, {Miglio}, {Deck},
  {Fabrycky}, {Montet}, {Buchhave}, {Chaplin}, {Hekker}, {Montalb{\'a}n},
  {Sanchis-Ojeda}, {Basu}, {Bedding}, {Campante}, {Christensen-Dalsgaard},
  {Elsworth}, {Stello}, {Arentoft}, {Ford}, {Gilliland}, {Handberg}, {Howard},
  {Isaacson}, {Johnson}, {Karoff}, {Kawaler}, {Kjeldsen}, {Latham}, {Lund},
  {Lundkvist}, {Marcy}, {Metcalfe}, {Silva Aguirre}, and
  {Winn}}]{2013Sci...342..331H}
{Huber} D, {Carter} JA, {Barbieri} M et~al. (2013) {Stellar Spin-Orbit
  Misalignment in a Multiplanet System}. Science 342:331--334

\bibitem[{{Irwin} and {Bouvier}(2009)}]{2009IAUS..258..363I}
{Irwin} J {Bouvier} J (2009) {The rotational evolution of low-mass stars}. In:
  {Mamajek} EE, {Soderblom} DR {Wyse} RFG (eds) The Ages of Stars, IAU
  Symposium, vol 258, pp 363--374, \doi{10.1017/S1743921309032025}

\bibitem[{{Jackson} et~al.(2008){Jackson}, {Greenberg}, and
  {Barnes}}]{2008ApJ...678.1396J}
{Jackson} B, {Greenberg} R {Barnes} R (2008) {Tidal Evolution of Close-in
  Extrasolar Planets}. \apj 678:1396-1406

\bibitem[{{Jackson} et~al.(2013){Jackson}, {Stark}, {Adams}, {Chambers}, and
  {Deming}}]{2013ApJ...779..165J}
{Jackson} B, {Stark} CC, {Adams} ER, {Chambers} J {Deming} D (2013) {A Survey
  for Very Short-period Planets in the Kepler Data}. \apj 779:165

\bibitem[{{Jackson} et~al.(2016){Jackson}, {Jensen}, {Peacock}, {Arras}, and
  {Penev}}]{2016CeMDA.126..227J}
{Jackson} B, {Jensen} E, {Peacock} S, {Arras} P {Penev} K (2016) {Tidal decay
  and stable Roche-lobe overflow of short-period gaseous exoplanets}. Celestial
  Mechanics and Dynamical Astronomy 126:227--248

\bibitem[{{Jackson} et~al.(2017){Jackson}, {Arras}, {Penev}, {Peacock}, and
  {Marchant}}]{2017ApJ...835..145J}
{Jackson} B, {Arras} P, {Penev} K, {Peacock} S {Marchant} P (2017) {A New Model
  of Roche Lobe Overflow for Short-period Gaseous Planets and Binary Stars}.
  \apj 835:145

\bibitem[{{Jia} and {Spruit}(2017)}]{2017MNRAS.465..149J}
{Jia} S {Spruit} HC (2017) {Instability of mass transfer in a planet-star
  system}. \mnras 465:149--160

\bibitem[{Jura(2003)}]{2003ApJ...584L..91J}
Jura M (2003) {A Tidally Disrupted Asteroid around the White Dwarf G29-38}. The
  Astrophysical Journal 584(2):L91--L94

\bibitem[{{Kawaler}(1988)}]{1988ApJ...333..236K}
{Kawaler} SD (1988) {Angular momentum loss in low-mass stars}. \apj
  333:236--247

\bibitem[{{Koenigl}(1991)}]{1991ApJ...370L..39K}
{Koenigl} A (1991) {Disk accretion onto magnetic T Tauri stars}. \apjl
  370:L39--L43

\bibitem[{{Kopal}(1959)}]{1959cbs..book.....K}
{Kopal} Z (1959) {Close binary systems}

\bibitem[{{Kunitomo} et~al.(2011){Kunitomo}, {Ikoma}, {Sato}, {Katsuta}, and
  {Ida}}]{2011ApJ...737...66K}
{Kunitomo} M, {Ikoma} M, {Sato} B, {Katsuta} Y {Ida} S (2011) {Planet
  Engulfment by \~{}1.5-3 M $_{sun}$ Red Giants}. \apj 737:66

\bibitem[{{Lai} et~al.(1993){Lai}, {Rasio}, and
  {Shapiro}}]{1993ApJS...88..205L}
{Lai} D, {Rasio} FA {Shapiro} SL (1993) {Ellipsoidal figures of equilibrium -
  Compressible models}. \apjs 88:205--252

\bibitem[{{Levrard} et~al.(2009){Levrard}, {Winisdoerffer}, and
  {Chabrier}}]{2009ApJ...692L...9L}
{Levrard} B, {Winisdoerffer} C {Chabrier} G (2009) {Falling Transiting
  Extrasolar Giant Planets}. \apjl 692:L9--L13

\bibitem[{{Lillo-Box} et~al.(2014){Lillo-Box}, {Barrado}, {Moya}, {Montesinos},
  {Montalb{\'a}n}, {Bayo}, {Barbieri}, {R{\'e}gulo}, {Mancini}, {Bouy}, and
  {Henning}}]{2014A&A...562A.109L}
{Lillo-Box} J, {Barrado} D, {Moya} A et~al. (2014) {Kepler-91b: a planet at the
  end of its life. Planet and giant host star properties via light-curve
  variations}. \aap 562:A109

\bibitem[{{Lin} and {Papaloizou}(1979)}]{1979MNRAS.186..799L}
{Lin} DNC {Papaloizou} J (1979) {Tidal torques on accretion discs in binary
  systems with extreme mass ratios}. \mnras 186:799--812

\bibitem[{{Lin} et~al.(1996){Lin}, {Bodenheimer}, and
  {Richardson}}]{1996Natur.380..606L}
{Lin} DNC, {Bodenheimer} P {Richardson} DC (1996) {Orbital migration of the
  planetary companion of 51 Pegasi to its present location}. \nat 380:606--607

\bibitem[{{Liu} et~al.(2013){Liu}, {Guillochon}, {Lin}, and
  {Ramirez-Ruiz}}]{2013ApJ...762...37L}
{Liu} SF, {Guillochon} J, {Lin} DNC {Ramirez-Ruiz} E (2013) {On the
  Survivability and Metamorphism of Tidally Disrupted Giant Planets: The Role
  of Dense Cores}. \apj 762:37

\bibitem[{Livio and Soker(1984)}]{1984MNRAS.208..763L}
Livio M Soker N (1984) {Star-planet systems as possible progenitors of
  cataclysmic binaries}. Monthly Notices of the Royal Astronomical Society
  (ISSN 0035-8711) 208:763--781

\bibitem[{{Mandell} et~al.(2007){Mandell}, {Raymond}, and
  {Sigurdsson}}]{2007ApJ...660..823M}
{Mandell} AM, {Raymond} SN {Sigurdsson} S (2007) {Formation of Earth-like
  Planets During and After Giant Planet Migration}. \apj 660:823--844

\bibitem[{{Mardling} and {Lin}(2002)}]{2002ApJ...573..829M}
{Mardling} RA {Lin} DNC (2002) {Calculating the Tidal, Spin, and Dynamical
  Evolution of Extrasolar Planetary Systems}. \apj 573:829--844

\bibitem[{{Matsumura} et~al.(2008){Matsumura}, {Takeda}, and
  {Rasio}}]{2008ApJ...686L..29M}
{Matsumura} S, {Takeda} G {Rasio} FA (2008) {On the Origins of Eccentric
  Close-In Planets}. \apjl 686:L29

\bibitem[{McQuillan et~al.(2013)McQuillan, Mazeh, and
  Aigrain}]{2013ApJ...775L..11M}
McQuillan A, Mazeh T Aigrain S (2013) {Stellar Rotation Periods of the Kepler
  Objects of Interest: A Dearth of Close-in Planets around Fast Rotators}. The
  Astrophysical Journal Letters 775(1):L11

\bibitem[{Mel{\'e}ndez et~al.(2009)Mel{\'e}ndez, Asplund, Gustafsson, and
  Yong}]{melendez09}
Mel{\'e}ndez J, Asplund M, Gustafsson B Yong D (2009) {THE PECULIAR SOLAR
  COMPOSITION AND ITS POSSIBLE RELATION TO PLANET FORMATION}. The Astrophysical
  Journal 704(1):L66--L70

\bibitem[{{Metzger} et~al.(2012){Metzger}, {Giannios}, and
  {Spiegel}}]{2012MNRAS.425.2778M}
{Metzger} BD, {Giannios} D {Spiegel} DS (2012) {Optical and X-ray transients
  from planet-star mergers}. \mnras 425:2778--2798

\bibitem[{{Murray} and {Dermott}(1999)}]{1999ssd..book.....M}
{Murray} CD {Dermott} SF (1999) {Solar system dynamics}

\bibitem[{{Mustill} et~al.(2015){Mustill}, {Davies}, and
  {Johansen}}]{2015ApJ...808...14M}
{Mustill} AJ, {Davies} MB {Johansen} A (2015) {The Destruction of Inner
  Planetary Systems during High-eccentricity Migration of Gas Giants}. \apj
  808:14

\bibitem[{{Nagasawa} et~al.(2008){Nagasawa}, {Ida}, and
  {Bessho}}]{2008ApJ...678..498N}
{Nagasawa} M, {Ida} S {Bessho} T (2008) {Formation of Hot Planets by a
  Combination of Planet Scattering, Tidal Circularization, and the Kozai
  Mechanism}. \apj 678:498-508

\bibitem[{{Naoz}(2016)}]{2016ARA&A..54..441N}
{Naoz} S (2016) {The Eccentric Kozai-Lidov Effect and Its Applications}. \araa
  54:441--489

\bibitem[{{Nordhaus} and {Blackman}(2006)}]{2006MNRAS.370.2004N}
{Nordhaus} J {Blackman} EG (2006) {Low-mass binary-induced outflows from
  asymptotic giant branch stars}. \mnras 370:2004--2012

\bibitem[{{Ogilvie}(2014)}]{2014ARA&A..52..171O}
{Ogilvie} GI (2014) {Tidal Dissipation in Stars and Giant Planets}. \araa
  52:171--210

\bibitem[{{Paczy{\'n}ski}(1971)}]{1971ARA&A...9..183P}
{Paczy{\'n}ski} B (1971) {Evolutionary Processes in Close Binary Systems}.
  \araa 9:183

\bibitem[{{Paxton} et~al.(2011){Paxton}, {Bildsten}, {Dotter}, {Herwig},
  {Lesaffre}, and {Timmes}}]{2011ApJS..192....3P}
{Paxton} B, {Bildsten} L, {Dotter} A et~al. (2011) {Modules for Experiments in
  Stellar Astrophysics (MESA)}. \\apjs 192:3,
  \urlprefix\url{http://dx.doi.org/10.1088/0067-0049/192/1/3}

\bibitem[{{Paxton} et~al.(2013){Paxton}, {Cantiello}, {Arras}, {Bildsten},
  {Brown}, {Dotter}, {Mankovich}, {Montgomery}, {Stello}, {Timmes}, and
  {Townsend}}]{2013ApJS..208....4P}
{Paxton} B, {Cantiello} M, {Arras} P et~al. (2013) {Modules for Experiments in
  Stellar Astrophysics (MESA): Planets, Oscillations, Rotation, and Massive
  Stars}. \\apjs 208:4,
  \urlprefix\url{http://dx.doi.org/10.1088/0067-0049/208/1/4}

\bibitem[{{Paxton} et~al.(2015){Paxton}, {Marchant}, {Schwab}, {Bauer},
  {Bildsten}, {Cantiello}, {Dessart}, {Farmer}, {Hu}, {Langer}, {Townsend},
  {Townsley}, and {Timmes}}]{2015ApJS..220...15P}
{Paxton} B, {Marchant} P, {Schwab} J et~al. (2015) {Modules for Experiments in
  Stellar Astrophysics (MESA): Binaries, Pulsations, and Explosions}. \apjs
  220:15

\bibitem[{{Penev} et~al.(2012){Penev}, {Jackson}, {Spada}, and
  {Thom}}]{2012ApJ...751...96P}
{Penev} K, {Jackson} B, {Spada} F {Thom} N (2012) {Constraining Tidal
  Dissipation in Stars from the Destruction Rates of Exoplanets}. \apj 751:96

\bibitem[{{Priedhorsky} and {Verbunt}(1988)}]{Priedhorsky1988Tidal}
{Priedhorsky} WC {Verbunt} F (1988) Tidal forces and mass transfer
  instabilities in low-mass x-ray binaries. Astrophys J 333:895--905,
  \urlprefix\url{http://dx.doi.org/10.1086/166798}

\bibitem[{Ram{\'\i}rez et~al.(2009)Ram{\'\i}rez, Mel{\'e}ndez, and
  Asplund}]{2009A&A...508L..17R}
Ram{\'\i}rez I, Mel{\'e}ndez J Asplund M (2009) {Accurate abundance patterns of
  solar twins and analogs. Does the anomalous solar chemical composition come
  from planet formation?} A{\&}A 508(1):L17--L20

\bibitem[{{Rappaport} et~al.(2013){Rappaport}, {Sanchis-Ojeda}, {Rogers},
  {Levine}, and {Winn}}]{2013ApJ...773L..15R}
{Rappaport} S, {Sanchis-Ojeda} R, {Rogers} LA, {Levine} A {Winn} JN (2013) {The
  Roche Limit for Close-orbiting Planets: Minimum Density, Composition
  Constraints, and Application to the 4.2 hr Planet KOI 1843.03}. \apjl 773:L15

\bibitem[{{Rasio} and {Ford}(1996)}]{1996Sci...274..954R}
{Rasio} FA {Ford} EB (1996) {Dynamical instabilities and the formation of
  extrasolar planetary systems}. Science 274:954--956

\bibitem[{{Richardson} et~al.(1998){Richardson}, {Bottke}, and
  {Love}}]{1998Icar..134...47R}
{Richardson} DC, {Bottke} WF {Love} SG (1998) {Tidal Distortion and Disruption
  of Earth-Crossing Asteroids}. \icarus 134:47--76

\bibitem[{Ritter(1988)}]{Ritter1988Turning}
Ritter H (1988) {Turning on and off mass transfer in cataclysmic binaries}.
  Astronomy and Astrophysics 202:93--100

\bibitem[{Sackmann and Boothroyd(1999)}]{1999ApJ...510..217S}
Sackmann IJ Boothroyd AI (1999) {Creation of $^7$Li and Destruction of $^3$He,
  $^9$Be, $^10$B, and $^11$B in Low-Mass Red Giants, Due to Deep Circulation}.
  The Astrophysical Journal 510(1):217--231

\bibitem[{{Sanchis-Ojeda} et~al.(2013){Sanchis-Ojeda}, {Rappaport}, {Winn},
  {Levine}, {Kotson}, {Latham}, and {Buchhave}}]{2013ApJ...774...54S}
{Sanchis-Ojeda} R, {Rappaport} S, {Winn} JN et~al. (2013) {Transits and
  Occultations of an Earth-sized Planet in an 8.5 hr Orbit}. \apj 774:54

\bibitem[{{Sanchis-Ojeda} et~al.(2014){Sanchis-Ojeda}, {Rappaport}, {Winn},
  {Kotson}, {Levine}, and {El Mellah}}]{2014ApJ...787...47S}
{Sanchis-Ojeda} R, {Rappaport} S, {Winn} JN et~al. (2014) {A Study of the
  Shortest-period Planets Found with Kepler}. \apj 787:47

\bibitem[{{Schlaufman} and {Winn}(2016)}]{2016ApJ...825...62S}
{Schlaufman} KC {Winn} JN (2016) {The Occurrence of Additional Giant Planets
  Inside the Water-Ice Line in Systems with Hot Jupiters: Evidence Against
  High-Eccentricity Migration}. \apj 825:62

\bibitem[{{Setiawan} et~al.(2010){Setiawan}, {Klement}, {Henning}, {Rix},
  {Rochau}, {Rodmann}, and {Schulze-Hartung}}]{2010Sci...330.1642S}
{Setiawan} J, {Klement} RJ, {Henning} T et~al. (2010) {A Giant Planet Around a
  Metal-Poor Star of Extragalactic Origin}. Science 330:1642

\bibitem[{{Showman} and {Guillot}(2002)}]{2002A&A...385..166S}
{Showman} AP {Guillot} T (2002) {Atmospheric circulation and tides of ``51
  Pegasus b-like'' planets}. \aap 385:166--180

\bibitem[{Siess and Livio(1999{\natexlab{a}})}]{1999MNRAS.304..925S}
Siess L Livio M (1999{\natexlab{a}}) {The accretion of brown dwarfs and planets
  by giant stars - I. Asymptotic giant branch stars}. MNRAS 304(4):925--937

\bibitem[{Siess and Livio(1999{\natexlab{b}})}]{1999MNRAS.308.1133S}
Siess L Livio M (1999{\natexlab{b}}) {The accretion of brown dwarfs and planets
  by giant stars - II. Solar-mass stars on the red giant branch}. MNRAS
  308(4):1133--1149

\bibitem[{{Skumanich}(1972)}]{1972ApJ...171..565S}
{Skumanich} A (1972) {Time Scales for CA II Emission Decay, Rotational Braking,
  and Lithium Depletion}. \apj 171:565

\bibitem[{{Stauffer} and {Hartmann}(1987)}]{1987ApJ...318..337S}
{Stauffer} JR {Hartmann} LW (1987) {The distribution of rotational velocities
  for low-mass stars in the Pleiades}. \apj 318:337--355

\bibitem[{{Steffen} et~al.(2012){Steffen}, {Ragozzine}, {Fabrycky}, {Carter},
  {Ford}, {Holman}, {Rowe}, {Welsh}, {Borucki}, {Boss}, {Ciardi}, and
  {Quinn}}]{2012PNAS..109.7982S}
{Steffen} JH, {Ragozzine} D, {Fabrycky} DC et~al. (2012) {Kepler constraints on
  planets near hot Jupiters}. Proceedings of the National Academy of Science
  109:7982--7987

\bibitem[{Teitler and K{\"o}nigl(2014)}]{2014ApJ...786..139T}
Teitler S K{\"o}nigl A (2014) {Why is there a Dearth of Close-in Planets around
  Fast-rotating Stars?} The Astrophysical Journal 786(2):139

\bibitem[{{Th{\'e}ado} and {Vauclair}(2012)}]{2012ApJ...744..123T}
{Th{\'e}ado} S {Vauclair} S (2012) {Metal-rich Accretion and Thermohaline
  Instabilities in Exoplanet-host Stars: Consequences on the Light Elements
  Abundances}. \apj 744:123

\bibitem[{{Trilling} et~al.(1998){Trilling}, {Benz}, {Guillot}, {Lunine},
  {Hubbard}, and {Burrows}}]{1998ApJ...500..428T}
{Trilling} DE, {Benz} W, {Guillot} T et~al. (1998) {Orbital Evolution and
  Migration of Giant Planets: Modeling Extrasolar Planets}. \apj 500:428--439

\bibitem[{{Tucci Maia} et~al.(2014){Tucci Maia}, {Mel{\'e}ndez}, and
  {Ram{\'{\i}}rez}}]{2014ApJ...790L..25T}
{Tucci Maia} M, {Mel{\'e}ndez} J {Ram{\'{\i}}rez} I (2014) {High Precision
  Abundances in the 16 Cyg Binary System: A Signature of the Rocky Core in the
  Giant Planet}. \apjl 790:L25

\bibitem[{{Valsecchi} and {Rasio}(2014)}]{2014ApJ...787L...9V}
{Valsecchi} F {Rasio} FA (2014) {Planets on the Edge}. \apjl 787:L9

\bibitem[{{Valsecchi} et~al.(2015){Valsecchi}, {Rappaport}, {Rasio},
  {Marchant}, and {Rogers}}]{2015ApJ...813..101V}
{Valsecchi} F, {Rappaport} S, {Rasio} FA, {Marchant} P {Rogers} LA (2015)
  {Tidally-driven Roche-lobe Overflow of Hot Jupiters with MESA}. \apj 813:101

\bibitem[{{van Eyken} et~al.(2012){van Eyken}, {Ciardi}, {von Braun}, {Kane},
  {Plavchan}, {Bender}, {Brown}, {Crepp}, {Fulton}, {Howard}, {Howell},
  {Mahadevan}, {Marcy}, {Shporer}, {Szkody}, {Akeson}, {Beichman}, {Boden},
  {Gelino}, {Hoard}, {Ram{\'{\i}}rez}, {Rebull}, {Stauffer}, {Bloom}, {Cenko},
  {Kasliwal}, {Kulkarni}, {Law}, {Nugent}, {Ofek}, {Poznanski}, {Quimby},
  {Walters}, {Grillmair}, {Laher}, {Levitan}, {Sesar}, and
  {Surace}}]{2012ApJ...755...42V}
{van Eyken} JC, {Ciardi} DR, {von Braun} K et~al. (2012) {The PTF Orion
  Project: A Possible Planet Transiting a T-Tauri Star}. \apj 755:42

\bibitem[{{Van Laerhoven} and {Greenberg}(2012)}]{2012CeMDA.113..215V}
{Van Laerhoven} C {Greenberg} R (2012) {Characterizing multi-planet systems
  with classical secular theory}. Celestial Mechanics and Dynamical Astronomy
  113:215--234

\bibitem[{{Veras} et~al.(2014){Veras}, {Leinhardt}, {Bonsor}, and
  {G{\"a}nsicke}}]{2014MNRAS.445.2244V}
{Veras} D, {Leinhardt} ZM, {Bonsor} A {G{\"a}nsicke} BT (2014) {Formation of
  planetary debris discs around white dwarfs - I. Tidal disruption of an
  extremely eccentric asteroid}. \mnras 445:2244--2255

\bibitem[{{Villaver} and {Livio}(2007)}]{2007ApJ...661.1192V}
{Villaver} E {Livio} M (2007) {Can Planets Survive Stellar Evolution?} \apj
  661:1192--1201

\bibitem[{{Villaver} and {Livio}(2009)}]{2009ApJ...705L..81V}
{Villaver} E {Livio} M (2009) {The Orbital Evolution of Gas Giant Planets
  Around Giant Stars}. \apjl 705:L81--L85

\bibitem[{Wallerstein and Sneden(1982)}]{1982ApJ...255..577W}
Wallerstein G Sneden C (1982) {A K giant with an unusually high abundance of
  lithium - HD 112127}. ApJ 255:577--584

\bibitem[{{Weiss} and {Marcy}(2014)}]{2014ApJ...783L...6W}
{Weiss} LM {Marcy} GW (2014) {The Mass-Radius Relation for 65 Exoplanets
  Smaller than 4 Earth Radii}. \apjl 783:L6

\bibitem[{{Winn} et~al.(2010){Winn}, {Fabrycky}, {Albrecht}, and
  {Johnson}}]{2010ApJ...718L.145W}
{Winn} JN, {Fabrycky} D, {Albrecht} S {Johnson} JA (2010) {Hot Stars with Hot
  Jupiters Have High Obliquities}. \apjl 718:L145--L149

\bibitem[{Zuckerman et~al.(2007)Zuckerman, Koester, Melis, Hansen, and
  Jura}]{2007ApJ...671..872Z}
Zuckerman B, Koester D, Melis C, Hansen BM Jura M (2007) {The Chemical
  Composition of an Extrasolar Minor Planet}. The Astrophysical Journal
  671(1):872--877

\end{thebibliography}

\end{document}